\documentclass{ws-procs9x6}

\newcommand{\refeq}[1]{(\ref{#1})}
\def\etal {{\it et al.}}

\begin{document}

\title{OBSERVER AND PARTICLE TRANSFORMATIONS AND NEWTON'S LAWS}

\author{T.H.\ BERTSCHINGER$^*$, NATASHA A.\ FLOWERS, and JAY D.\ TASSON}

\address{Department of Physics and Astronomy, Carleton College\\
Northfield, MN 55057, USA\\
$^*$E-mail: bertschingert@carleton.edu}

\begin{abstract}
A frequently confused point in studies of symmetry violation is the distinction between observer and particle transformations. In this work, we consider a model in which a coefficient in the Standard-Model Extension leads to violations of rotation invariance in Newton's second law. The model highlights the distinction between observer and particle transformations. 
\end{abstract}

\bodymatter

\section{Introduction}

The Standard-Model Extension (SME) provides a general field-theoretic framework for studying Lorentz violation\cite{kostelec}, including rotation-invariance violation. To highlight the basic ideas of Lorentz-symmetry breaking, we consider rotation-invariance violation in Newton's second law:\cite{bert}
\begin{equation}
\label{mass}
F_j = m_{jk}a_k.
\end{equation}
Here $m_{jk}$ is a symmetric direction-dependent inertial mass (we consider conventional gravitational mass). This yields a valid and more general form that Newton himself could have chosen. 

Our effective inertial mass can be generated as a low-energy limit of the SME:\cite{kostelec11}
\begin{equation}
m_{jk} = m\left(\delta_{jk} + 2c_{jk}\right).
\end{equation}
Here $c_{jk}$ is a coefficient for Lorentz violation found in the fermion sector,
taken as symmetric.
It is depicted with background diamonds in Fig.\ 1.
A similar construction can also be found
associated with other SME coefficients for Lorentz violation. \cite{da}

Using a block on an inclined plane, we show that observer rotation invariance holds, while a particle rotation changes the experiment's outcome, violating Lorentz symmetry. 

\section{Block on an inclined plane}

Examine a block on an inclined plane devoid of friction. The $x$-axis points down the plane while the $y$-axis is perpendicular to the surface as shown in the left-hand diagram of Fig.\ 1. Let the block's effective inertial mass be diagonal: 
\begin{equation}
m_{jk} = m\left(\begin{array}{ccc}
1  +2c_{xx} & 0 & 0 \\
0 & 1 + 2c_{yy} & 0 \\
0 & 0 & 1 + 2c_{zz}
\end{array}\right).
\end{equation}
Solving for the particle's acceleration under the constraint $a_y = 0$ yields
\begin{equation}
a_x = a_R = (1 - 2c_{xx})g\sin\theta + O(c^2),
\label{eq1}
\end{equation}
where $a_R$ is the acceleration down the ramp. The only difference here from the conventional problem is the presence of $c_{xx}$. Qualitatively, the motion is down the plane with constant acceleration, as in the absence of Lorentz violation. 

\section{Observer-rotation invariance remains}

Perform an observer rotation on the original experiment; 
that is, consider the same problem in new coordinates
as shown in the left-hand diagram of Fig.\ 1. Here,
\begin{equation}
m_{j'k'} = m\left(\begin{array}{ccc}
1 + 2c_{x'x'} & 2c_{x'y'} & 0 \\
2c_{x'y'} & 1 + 2c_{y'y'} & 0 \\
0 & 0 & 1 + 2c_{z'z'}
\end{array}\right)
\end{equation}
is the mass in the new coordinates
obtained via $m'=RmR^T$, 
$R$ being a rotation matrix. Components are related by, for example,  
\begin{equation}
c_{x'x'} = c_{xx}\cos^2\theta + c_{yy}\sin^2\theta.
\end{equation}
Again solving, we find up to $O(c^2)$ the components 
\begin{eqnarray}
a_{x'} & = & (1 -2 c_{x'x'} \cos^2 \theta -2 c_{y'y'} \sin^2 \theta + 4c_{x'y'}\sin\theta\cos\theta)g\sin\theta\cos\theta,
%+O(c^2),
\nonumber \\ 
a_{y'} & = & -(1 -2 c_{x'x'} \cos^2 \theta -2 c_{y'y'} \sin^2 \theta + 4c_{x'y'}\sin\theta\cos\theta)g\sin^2\theta.
%+O(c^2).
\nonumber \\
\end{eqnarray}
However,
the acceleration is again
along the ramp with the magnitude found in  Eq.\ \refeq{eq1},
which is obtained from $a_{x'}$ and $a_{y'}$ by $a_R=\sqrt{a_{x'}^2+a_{y'}^2}$
and a substitution for $c_{j'k'}$ in terms of $c_{jk}$. 
Hence,
both observers agree on the outcome of the experiment.

\begin{figure}[t]
\psfig{file=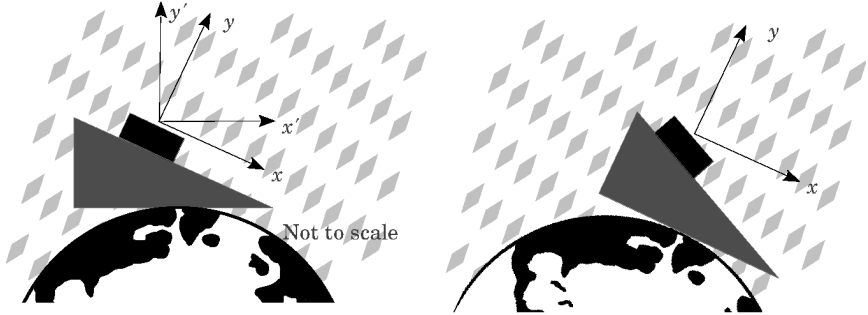,width=\textwidth}
\vskip-40pt
\caption{Diagrams.}
\end{figure}

\section{Particle-rotation invariance is violated}

A particle rotation
of the original system
leaves the mass matrix as in Eq.\ \eqref{mass},
but alters the direction of the gravitational field
relative to the background as shown
in the right-hand diagram of Fig.\ 1.
This produces an observably different acceleration. 
Solving for the motion of the particle subject to the constraint yields
\begin{eqnarray}
a_y & = & -(1 - 2c_{xx} - 2c_{yy})g\sin^2\theta + O(c^2),
\nonumber \\
a_x & = & (1 - 2c_{xx} - 2c_{yy})g\sin\theta\cos\theta + O(c^2). 
\end{eqnarray}
Here the component along the ramp is 
\begin{equation}
a_R = (1 - 2c_{xx} - 2c_{yy})g\sin\theta + O(c^2),
\end{equation}
which is different from the first cases, revealing observable Lorentz violation.

\end{document}